\documentclass[twocolumn,12pt]{article}

\usepackage{graphicx,amsfonts,amssymb,amsmath}
\usepackage[pagebackref=true,colorlinks,linkcolor=blue,citecolor=red]{hyperref}
\usepackage{subfigure,epstopdf,color,amsthm}

\usepackage{setspace}
\UseRawInputEncoding 
\usepackage[square,numbers,sort&compress]{natbib}
\usepackage{lineno}
\usepackage[font=footnotesize,skip=2pt]{caption}
\usepackage{lineno}
\onehalfspace

\begin{document}
\onecolumn

\begin{center}
{\large\bf Effect of Interfacial Defects on the Electronic Properties of MoS$_2$ Based Lateral T-H Heterophase Junctions}

Mohammad Bahmani$^{1,\dag}$, Mahdi Ghorbani-Asl$^{2,\ddag}$, and Thomas Frauenheim$^{\star, 1,3,4}$

{\it $^1$ Bremen Center for Computational Materials Science (BCCMS), Department of Physics, Bremen University, 28359 Bremen, Germany}

{\it $^2$ Institute of Ion Beam Physics and Materials Research, Helmholtz-Zentrum Dresden-Rossendorf, 01328 Dresden, Germany}

{\it $^3$ Beijing Computational Science Research Center (CSRC), 100193 Beijing, China}
 
{\it $^4$ Shenzhen JL Computational Science and Applied Research Institute, 518110 Shenzhen, China}
\end{center}

\bigskip

The coexistence of semiconducting (2H) and metallic (1T) phases of MoS$_{2}$ monolayers have further pushed their strong potential for applications in the next generation of electronic devices based on the two-dimensional lateral heterojunctions. Structural defects have considerable effects on the properties of these 2D devices. In particular, the interfaces of two phases are often imperfect and may contain numerous vacancies created by phase engineering techniques, e.g. under the electron beam.
Here, the transport behaviors of the heterojunctions in the existence of point defects are explored by means of first-principles calculations and non-equilibrium Green's function approach.
While vacancies in semiconducting MoS$_{2}$ act as scattering centers, their presence at the interface improves the flow of the charge carriers. In the case of $V_{Mo}$, the current has been increased by two orders of magnitude in comparison to the perfect device. The enhancement of transmission was explained by changes in the electronic densities at the T-H interface, which open new transport channels for electron conduction.

%\bigskip
%
%\noindent PACS numbers(s):?????????????
%\bigskip

\noindent\rule{5truecm}{0.01in}\\
\noindent $^\dag$ mbahmani$@$uni-bremen.de, 
$^\ddag$ mahdi.ghorbani$@$hzdr.de,
$^\star$ frauenheim$@$uni-bremen.de 
%\maketitle %\baselineskip 24pt

%\twocolumn
\newpage
%---------------------------------------------
% Introduction
% ---------------------------------------------
\section{Introduction}
Among the developing family of two-dimensional (2D) materials, transition metal dichalcogenides (TMDs) provide one of the most diverse electronic properties including topological insulators, semiconductors, (semi)metals and superconductors \cite{Avalos-Ovando2019,Wang2019,Houssa2020}. Noticeably, such a difference in the electronic structure of TMDs correlates with their structural configurations, called phases \cite{Chhowalla2013}. Monolayers of MoS$_{2}$ in H-phase, with trigonal prismatic coordination of metal atoms is a semiconducting material \cite{Nourbakhsh2016,Marian2017a}, while T-phase with octahedral coordination shows metallic character. The H-phase monolayer is reported to be a promising material for field-effect transistors (FETs) with small-scale channel lengths and negligible current leakage \cite{Nourbakhsh2016,Marian2017a}. 

Recent experiments have already shown controlled transitions from one phase to another via external stimuli such as electron beam \cite{Lin2014},  ion intercalation \cite{Kappera2014}, or laser irradiation \cite{Cho2015}. These phase-engineered 2D materials with minimum variations in atomic structure and uniform stoichiometry not only demonstrate rich physical behavior but also open up new avenues for the design of electronic devices. 
The fabrication of lateral metallic/semiconducting heterostructures has been suggested as a practical method to minimize the contact resistance at the interface between 2D semiconductors and metal electrodes. In particular, the formation of covalent bonds between the two phases can introduce paths for carriers to travel across the interfaces, thus, the Schottky barrier and contact resistance are reduced \cite{Kang2014,Aierken2018,Fan2018,Zhang2019}. It has also been demonstrated that 1T-phase engineered electrodes in MoS$_2$ based electronic devices would generate ohmic contacts and, as a result, improve electrical characteristics \cite{Friedman2016,Fan2018}.

Apart from intrinsic defects, the local phase transitions induced by electron beam irradiation may give rise to the formation of point defects, in particular at the interface of the two phases. \cite{Komsa2013,Yang2016,Han2016,Ghorbani-Asl2017,Ma2017,Klein2019,Barthelmi2020,Mitterreiter2020}.
Defects can also be intentionally introduced during the post-growth stage via ion bombardment, plasma treatment, vacuum annealing, or chemical etching.\cite{Komsa2013,Yang2016,Han2016,Ghorbani-Asl2017,Ma2017,Klein2019,Barthelmi2020,Mitterreiter2020}.
Indeed, the theoretical and experimental results showed that the presence of sulfur vacancies can decrease the energy difference between the H and T phases and eventually stabilize the 1T phase in MoS$_2$ monolayer \cite{Kretschmer2017,JZhu2017}.
The presence of point defects in semiconducting MoS$_2$ monolayers leads to the observation of the localized states in their electronic structure, which act as short-ranged scattering centers for charge carriers \cite{Ghorbani-Asl2013A,Pandey2016,Lin2016,Bahmani2020}. Hence, defects were found to deteriorate the mobility of the fabricated devices \cite{Novoselov2005,Radisavljevic2011,Lee2012}. It was also shown that sulfur line vacancies in MoS$_2$ can behave like pseudo-ballistic wire for electron transport \cite{Yong2008}.

So far, several theoretical studies have reported the transport properties of phase-engineered devices based on TMDs monolayers including MoS$_2$ based lateral junctions \cite{Hu2015,Sivaraman2016,Fan2017,Marian2017,Aierken2018,Fan2018}.
In most of these studies, however, it is assumed that two phases have a perfect crystalline structure and connected via an atomically sharp and defect-free interface.

%point defects are always present in the laboratory samples due to the kinetics of processing as well as the thermal equilibrium, which impose significant effects on their optical, magnetic, and especially electrical properties \cite{Ghorbani-Asl2013A,Tongay2013,Zhou2013,Feng2015,Pandey2016,Lin2016}.
 %For example in STEM experiment, the electron beams have been shown to acquire enough charge to induce S-plane gliding in the 2H-to-1T transition in MLs~MoS$_2$ and trigger the transformation \cite{Lin2014,Kappera2014,Wang2014,Zhao2017}. 
 
% Hence, it is most likely that phase-engineered devices based on 2D TMDs contain vacancies and antisites, in particular at the interface of the two phases. 

Here, transport properties of devices based on MLs~MoS$_2$, containing various point vacancies and antisites at the interface between metallic and semiconducting phases, are the subject of the present study. Our systematic investigations show significant improvements in the current, as molybdenum vacancy and vacancy complexes are created at the interfaces of two phases. 
%There are fundamental differences in transport characteristics of phase-engineered devices with and without defects. 
These findings render defect engineering as an efficient route to further improve the performance of the devices based on the lateral heterojunctions formed from TMDs.

%In the next section, the computational methods as well as the chosen parameters are explained, while section \ref{sec:resdis} contains a detailed discussion on the electronic and transport properties of the phase-engineered devices. In section \ref{sec:con}, we conclude our results and hypothesis.

%---------------------------------------------
% Computational details
% ---------------------------------------------
\section{Computational Details} \label{sec:compdetails}
Density-functional theory (DFT) calculations were performed using numerical atomic orbitals (NAOs) basis sets as implemented in SIESTA code \cite{Ordejon1996,Soler2002}. The norm-conserving pseudopotentials, including the effect of core electrons, are employed, which were obtained using the Troullier-Martin method \cite{Troullier1991A,Troullier1991B}. The Perdew-Burke-Ernzerhof (PBE) functional in the generalized gradient approximation (GGA) is used to describe the exchange and correlation interactions \cite{Perdew1996}.

%The wave function is expanded by a single-zeta basis set with one polarization function (SZP) to optimize the metal-semiconductor interfaces between 1T and 2H phases of MoS$_2$.
%The wave function is expanded by a double-zeta basis set with one polarization function (DZP) and 4p diffusive orbitals to optimize the metal-semiconductor interfaces between 1T and 2H phases of MoS$_2$.
%\textbf{Similar results have been obtained when the basis sets were enlarged.}
In the optimization calculations, the Brillouin zone (BZ) of supercells was sampled using a 9$\times$1$\times$3 Monkhorst–Pack grid. In the electronic and transport calculations, 5 and 29 k-points were used, respectively, along the transverse direction. The conjugate-gradients (CG) method was applied to optimize the lattice vectors and atomic positions of all the structures and interfaces. The geometries were considered relaxed when the Hellman-Feynman forces on each atom became smaller than 10 meV/\AA. The energy cut-off of 450 Ry is used in the framework of the real-space grid techniques to obtain Hartree, exchange, and correlation energies. The Split-Norm was set to 0.16 and the Energy-Shift of 0.02 Ry was chosen to determine the confinement radii. 
%\textbf{The same technique was used to find the equilibrium atomic position around defects at the interfaces and in the channel region.} 
The total energy convergency criteria ($\bigtriangleup E_{tot}$) is chosen to be $10^{-5}$eV for k-points and $10^{-4}$eV for energy cut-off.
When the difference between two consecutive steps was less than $10^{-4}$eV, the total energies in self-consistent field (SCF) cycles were considered converging.

The electron transport calculations were performed using non-equilibrium Green’s functions (NEGF) techniques, as implemented in TranSIESTA and TBtrans \cite{Brandbyge2002,Papior2017}. The same basis sets as for the electronic calculations, namely SZP, were employed for the transport calculations.
%similar to the previous simulations \cite{Brandbyge2002,Topsakal2010,Houssa2019}. 
The current through the heterophase junction under a finite bias voltage was calculated within the Landauer formula \cite{Landauer1970}

\begin{equation}
I = \frac{2 e}{h}\int Trace\left[G_{C}^{\dagger}(E) \Gamma_R (E) G_{C}(E) \Gamma_L (E) \right] \left( f_{L}(E) - f_{R}(E) \right) dE.
\end{equation}

%Here, parameters $e$ and $h$ are electron charge and the Plank constant, respectively.
Where $G_{C}^{\dagger}(E)$ and $G_{C}(E)$ are the retarded and advanced Green's functions of the channel region. The effect of left (L) and right (R) electrodes are projected onto the scattering region via their corresponding self-energies, $\Gamma_L (E)$ and $\Gamma_R (E)$. The Fermi distribution of $f_{L}(E)$ and $f_{R}(E)$ represent the available states for electrons in the left and right electrodes. The transport calculations were performed at 300 K.

%---------------------------------------------
% Results and discussion
% ---------------------------------------------
\section{Results and Discussion} \label{sec:resdis}
\begin{figure*}[!htb]
\centering
\includegraphics[width=1.0\textwidth]{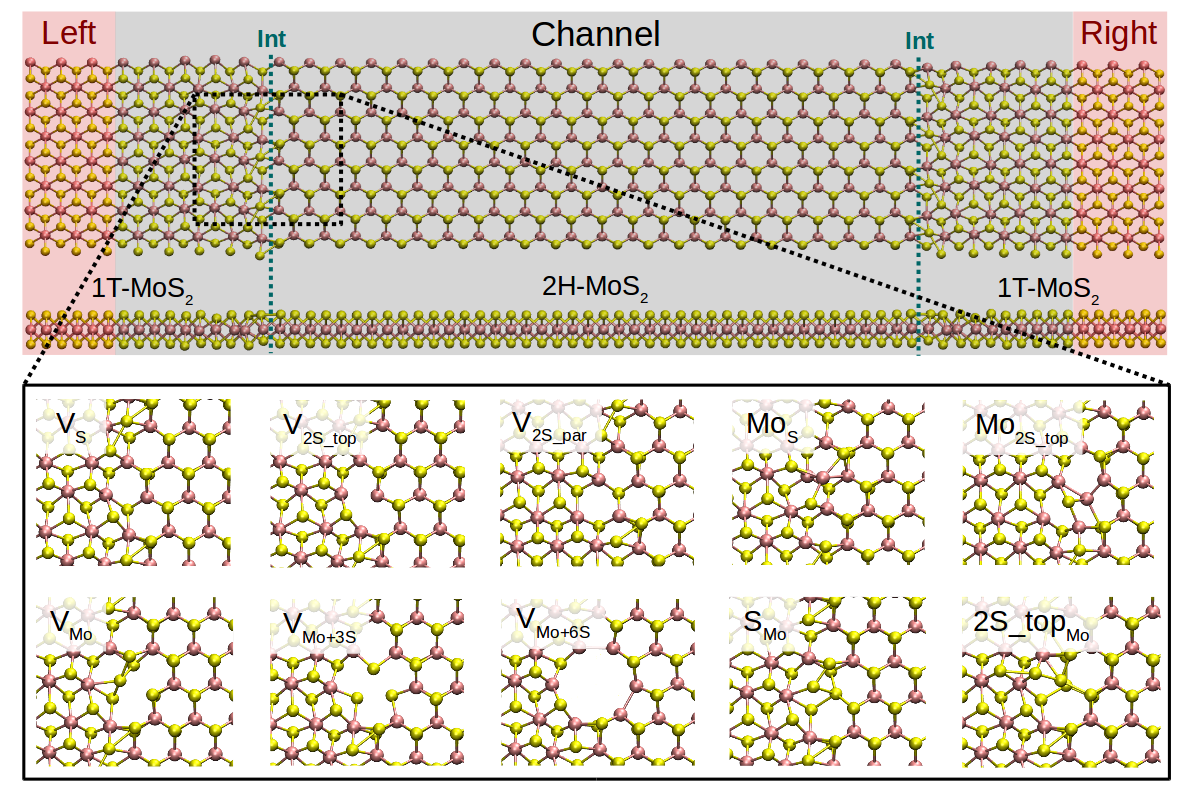}
\caption{Upper: Schematic of the T-H heterophase junction of MoS$_2$ monolayer. Electrodes (only 1T-MoS$_{2}$) and channel region (a combination of 1T- and 2H-MoS$_{2}$) are highlighted with shaded red and black, respectively. The interfaces are indicated with green dashed-lines. Lower: Optimized structures of the point defects at the left interface of the devices are shown. The complete defective devices are shown in Fig. S1 in the supplementary information.}
\label{fgr:schem}
\end{figure*}

Depending on the edge orientation of monolayers, armchair and zigzag interfaces can be realized. The armchair interfaces have been shown to be most stable against buckling \cite{Sivaraman2016,Aierken2018}. They are also energetically more favorable than connecting the zigzag terminated edges in the sulfur-rich limit \cite{Hu2015,Sivaraman2016}. The recent theoretical study showed that the conductivity of the armchair edges is higher than the zigzag interfaces due to the presence of metallic Mo zigzag chains along the transport direction \cite{Aierken2018}. Accordingly, we consider the armchair interface in the present study. In order to create Schottky contacts at the interfaces, the semiconducting 2H-phase of MoS$_{2}$ (channel region) is sandwiched between two metal electrodes of 1T-MoS$_{2}$, as shown in Fig. \ref{fgr:schem}.

The size of the whole device is 117.50 \AA\; along the transport direction (Z axis) and 22.00 \AA\; in the transverse direction (X axis), including the channel with a length of 98.46 \AA\; corresponding to 31 unit cells of MoS$_{2}$. The channel length is long enough to avoid artificial interactions between the two electrodes. Also, it includes small adjacent portions of the 1T phase as buffer layers to provide a computationally convenient configuration for calculating self-energies at the boundaries.\cite{Ghorbani-2015}. The periodic boundary conditions were applied along the axis transverse to the transport direction. A vacuum layer of 50 \AA\;normal to the monolayers was considered, which prevents interactions between adjacent supercells.
%In order to preserve the bulk-like characteristics of the electrodes, 5 layers of 1T-MoS$_{2}$ are added to both sides of the semiconducting region, which itself is made of 21 layers of 2H-MoS$_{2}$. Thus, the effect of interfaces on the states inside the electrodes are avoided. These 31 layers sum up to a channel length of 98.46 \AA.

Several junctions composed of 1T and 2H of MLs~MoS$_{2}$ are considered without defects (perfect) and when containing point defects in the phase boundaries, as shown in Fig. \ref{fgr:schem}.
%In Fig. \ref{fgr:schem}, optimized structures of only the left interface are shown, where the position of defects are denoted with blue arrows. 
%\textbf{The complete defective devices are displayed in Fig. S2 in the supplementary information.} 
We considered 10 types of point defects, most of which were observed in experiments \citep{Zhou2013,Yang2016,Han2016,Ma2017,Stanford2019,Klein2019} and their stability were analyzed by DFT calculations \citep{Komsa2013,Komsa2015,Lin2016,Bahmani2020}. 
We look at a sulfur and a molybdenum vacancy, $V_{S}$ and $V_{Mo}$, a double sulfur vacancy from upper and bottom layers $V_{2S-top}$, and the case of removing two atoms from the upper sulfur layer and parallel to the interface, $V_{2S-par}$. Besides, vacancy complexes of molybdenum and three sulfurs ($V_{Mo+3S}$) and six sulfurs ($V_{Mo+6S}$) are also studied. Four antisites are also considered: ${Mo}_{S}$, ${Mo}_{2S-top}$, ${S}_{Mo}$, and ${2S-top}_{Mo}$. 
%The stability of these defects in MoS$_{2}$ monolayers has been investigated in the previous theoretical reports. \citep{Komsa2013,Komsa2015,Lin2016,Bahmani2020}.  
Our electronic structure calculations indicate that 2H-MoS$_2$ is a semiconductor with a bandgap of 1.73 eV, while 1T-MoS$_2$ has a metallic character (see Fig. S2). These results are in agreement with previous theoretical reports at the same level of theory \cite{Roldan2014,Komsa2015,Saha2016a,Marian2017a,Sharma2018a}. Such studies have shown no difference or a difference of $\%0.63$ between the lattice constant of the T-~and~H-~phase of MoS$_2$ monolayers. Therefore, the same lattice parameters, namely 3.176\AA, are used for both phases. Such a phase transition can be seen as the collective displacement of sulfur atoms while the stoichiometry of the materials is preserved.
%one of the sulfur atoms are moved toward the center of the lattice to make the 1T-MoS$_2$ structure. 
The constructed lateral heterostructures with armchair edges are optimized, as shown at the top of Fig. \ref{fgr:schem}. It should be noted that the optimization could not transform 1T into the 2H phase but induce some distortions, indicating the activation barrier for the phase transformation is higher than the relaxation of the boundary. In addition, the atomic network can be subjected to strain as a result of defects in the phase boundary. Fig. S3 shows the strain map, which is specified as the total displacements in all three axes as compared to the perfect interface. It can be seen that the largest change in the atomic positions occurs in the case of $V_{Mo+3S}$ at the interfaces while the sulfur vacancies induce the smallest displacements into the phase boundary.

%Accordingly, the modifications due to the local strain are also included into their potential profile. 
%\textbf{In Fig. S3 in supplementary information, the calculated amount of displacement due to the absence of different combination of atoms are shown.} 
%Here, the displacements in all three axis are summed up to obtain the total movement of atoms. 

%Since synthesized samples always contain crystalline defects, we systematically investigate the effect of various point vacancies and antisites, located at both interfaces, on the device properties, as shown in Fig. \ref{fgr:schem}. Here, we analyze the change in current through defective devices based on transmission curves, local density of states (LDOS), and vector currents.
%Here, blue arrows point toward the position of the defects only at the left interface. 

%\newpage
% ---------------------------------------------
% ---------------------------------------------
\subsection{Sulfur vacancies}
\begin{figure}[!htb]
\centering
\includegraphics[width=0.85\textwidth]{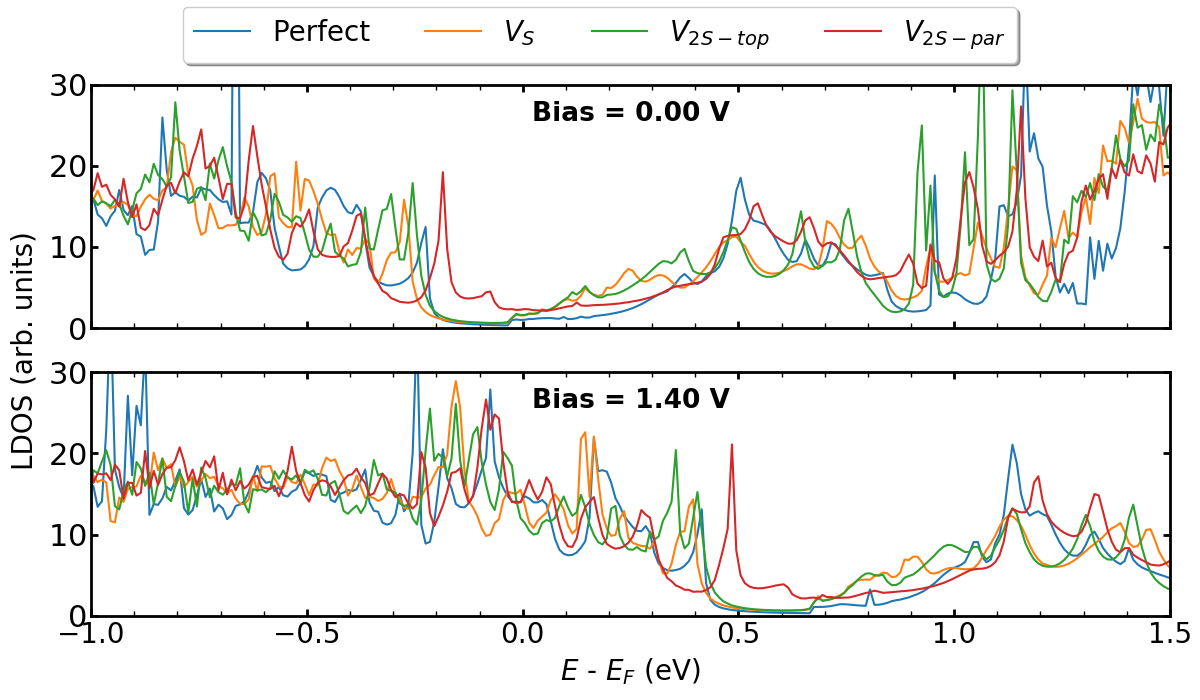}
\caption{(Color online) LDOS at the left interface of T-H heterophase junction containing various sulfur vacancies; $V_{S}$, $V_{2S-top}$, and $V_{2S-par}$. Energies are shifted with respect to their corresponding Fermi energy.}

\label{fgr:ldos_v_s}
\end{figure}

In this section, we present the electronic and transport properties of T-H heterophase junction containing interfacial sulfur vacancies; $V_{S}$, $V_{2S-top}$, and $V_{2S-par}$. In Fig. \ref{fgr:ldos_v_s}, local density of states (LDOS) on the atoms at the left interface of such devices are plotted at Bias $= 0.00 \,V$ and Bias $= 1.40 \,V$. 
In the following, the term "interface" is used for a part of the device, which consists of atoms from one layer of 1T-MoS$_2$ and one layer of 2H-MoS$_2$. Due to the electronic states from T phase, the band gap in LDOS is narrower than that for pristine 2H phase of MoS$_2$. Comparing the two figures, there is a shift in the energy, corresponding to half of the applied voltage.
%This is due to the fact that in transport code, bias is divided and assigned equally to each electrode, but with opposite signs. 

The presence of defects introduces new states close to the Fermi level and increases electron density at the interface, which is mainly contributed by metal $d$ orbitals. It is evident that defect-associated states are more pronounced in the case of V$_{2s-par}$ where the electron density is enhanced in the vicinity of the Fermi level, including a peak at -0.2 eV. The results showed that other types of sulfur vacancies have a negligible impact on the electronic structure around the Fermi energy. It should be noted that the sulfur vacancies in 2H-MoS$_2$ monolayers act as scattering centers and consequently diminish the transport properties \cite{Ghorbani-Asl2013A,Pandey2016,Lin2016,Bahmani2020}.

%. These defect states are almost situated
%under the bottom of the conduction band and completely
%contributed by metal d orbitals. On comparing Fig. 6(a) with Fig. 6(b), we found that the
%position of in-gap states with respect to the bulk band is almost
%unchanged while the energy splitting at the G-point increases
%when the S vacancy is approaching the interface. 
%It should be noted that in-gap states induced by S vacancy in the WS2 side are %much closer to the conduction band than those induced
%by S vacancy in the MoS2 side. As the S vacancy is located at the interface,
%both interfacing Mo and W atoms are not saturated. As can be
%seen from Fig. 6(c), the lower in-gap state (magenta curve) is
%attributed to Mo d orbitals while the upper one (green curve) is
%contributed by hybridized states of Mo and W. In comparison
%with the upper curve in Fig. 6(d) or (e), hybridized states of
%Mo and W shift far from the conduction band. As a result,
%in-gap states induced by S vacancy near the bottom of the
%conduction band for in-plane heterostructures of MoS2 and

%States, which are originated from the defects, have been induced within the bandgap. 

\begin{figure}[!htb]
\centering
\includegraphics[width=1.00\textwidth]{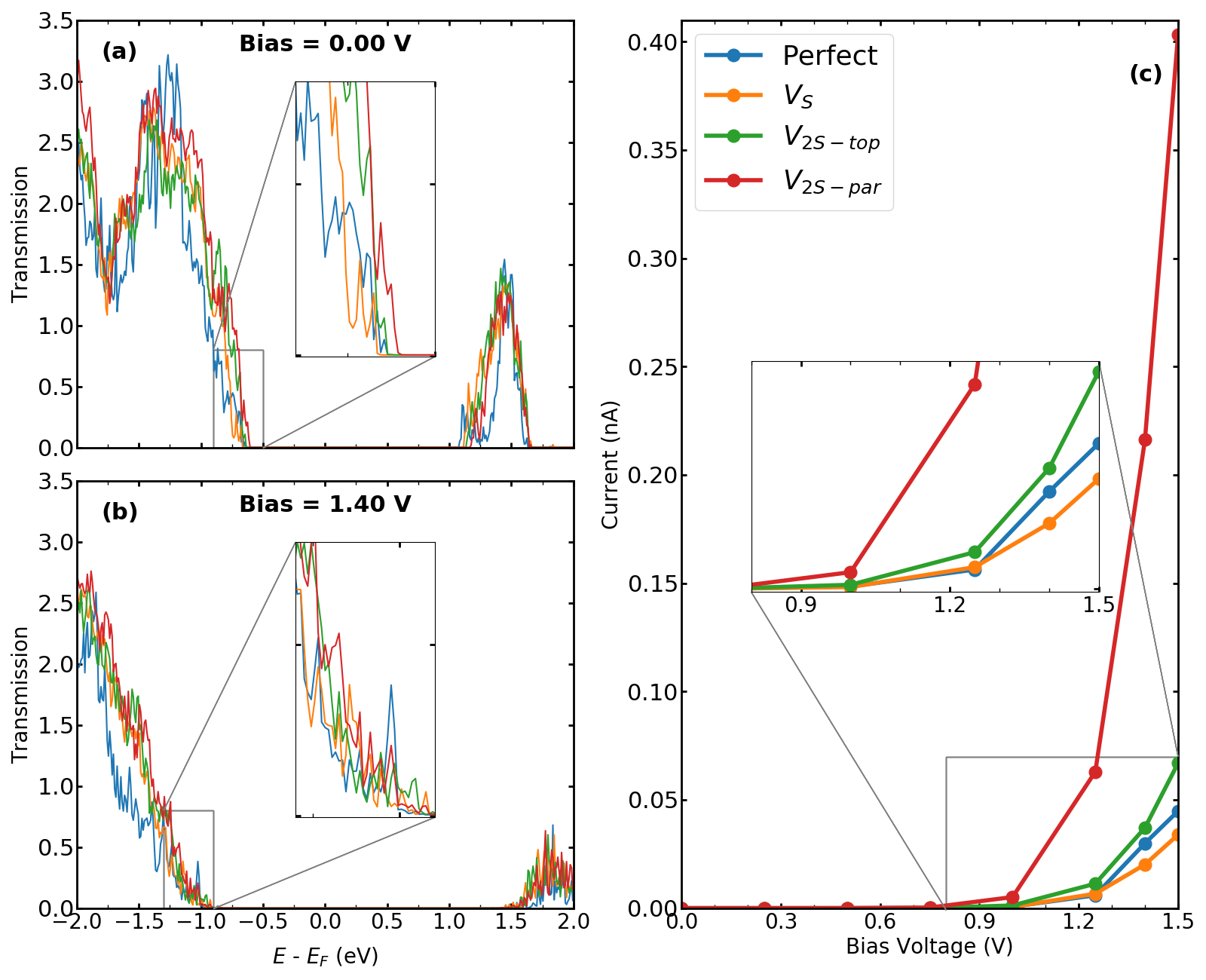}
\caption{
(Color online) Transmission spectra for T-H heterophase junction of MoS$_2$ monolayer containing various sulfur vacancies at both interfaces at a) Bias $= 0.00 \,V$ and b) Bias $= 1.40 \,V$. Energies are shifted with respect to their corresponding Fermi energy. The insets show the change in the electronic transmission channels at the top of the valance band. c) I-V characteristics for the same devices. The inset shows the current around the threshold voltage.}
\label{fgr:IV_trans_s}
\end{figure}

In order to elaborate the electron conduction dependency on the geometry of contact between the T and the H phases, transmission spectra for the junction without and with interfacial defects at two Bias, $0.00$ and $1.40 \,V$, are shown in Fig. \ref{fgr:IV_trans_s}a and \ref{fgr:IV_trans_s}b, respectively. Corresponding to the band gap of 2H-MoS$_2$, there is no transmission at zero bias within an energy range of $1.7 \,eV$ around the Fermi level. A comparison between perfect interface and those containing sulfur vacancies shows a growth in transmission probability, suggesting the contribution of defect states in electrical transport. Specifically, the transmission coefficients close to the valance band edge can be increased to almost two times for the case of $V_{2S-par}$ vacancy. 
The IV characteristics of the studied T-H heterophase junction are shown in Fig. \ref{fgr:IV_trans_s}c. The junction displays a non-linear current-voltage similar to the characteristics of a resonant tunneling diode. The energy mismatch between the Fermi energy of the metallic 1T electrodes and the lowest unoccupied levels of the 2H phase causes the presence of zero current and the need for threshold voltage to produce finite current flow through the junction. The value of threshold voltage was reduced from ~1.0 V for the perfect interface to ~0.75V for the interface with divacancy. The appearance of defect-associated resonant states in the transmission spectra within the voltage window changes the current through the system, leading to an increase by an order of magnitude, when $V_{2S-par}$ vacancy is present at the interfaces, as shown in Fig. \ref{fgr:IV_trans_s}c. 

% --------------------------------------------------------
% --------------------------------------------------------
\subsection{Molybdenum vacancies and vacancy complexes}
\begin{figure}[!htb]
\centering
\includegraphics[width=0.85\textwidth]{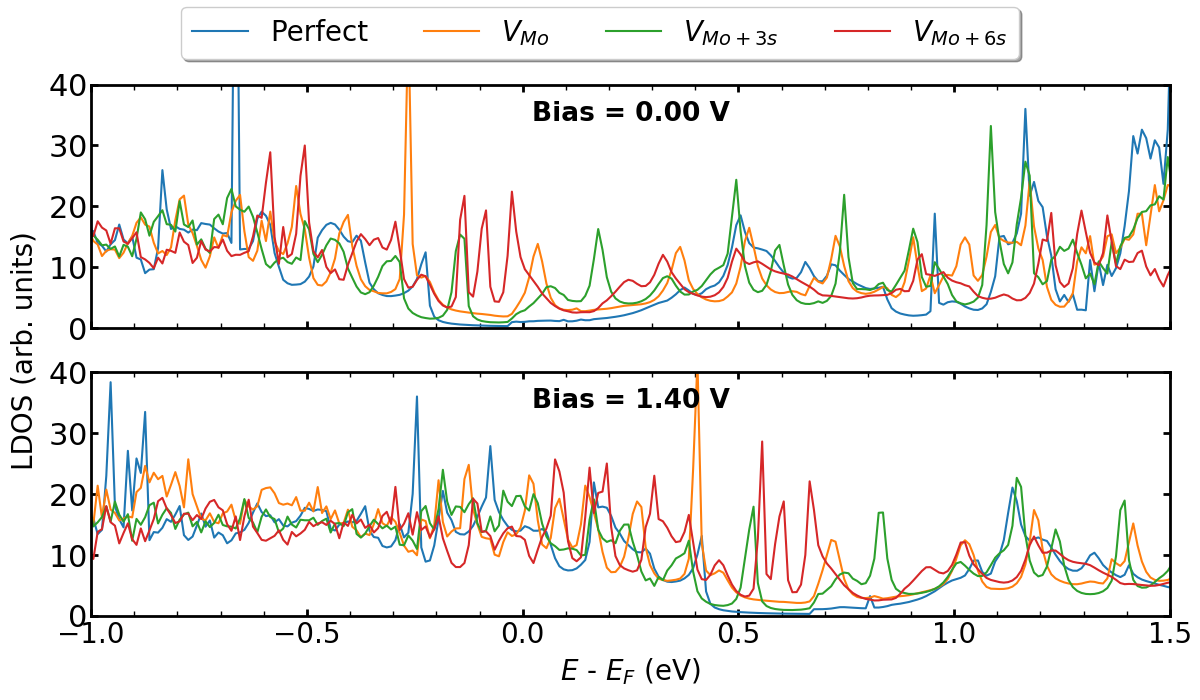}
\caption{(Color online) LDOS at the left interface of T-H heterophase junction containing molybdenum vacancy, $V_{Mo}$, and vacancy complexes as $V_{Mo+3S}$ and $V_{Mo+6S}$. Energies are shifted with respect to their corresponding Fermi energy.}
\label{fgr:ldos_v_mo}
\end{figure}

\begin{figure}[!htb]
\centering
\includegraphics[width=1.00\textwidth]{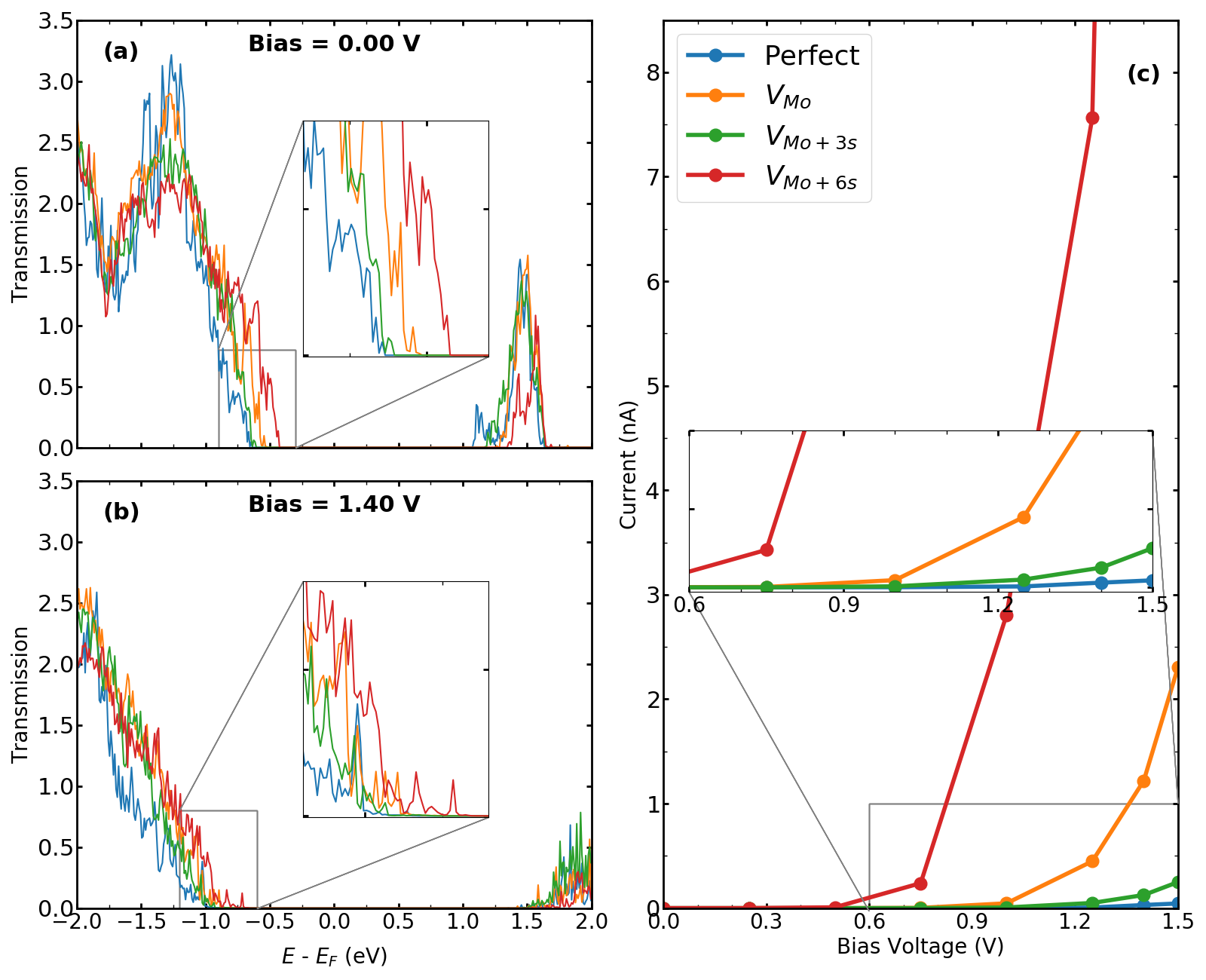}
\caption{(Color online) Transmission spectra for T-H heterophase junction of MoS$_2$ monolayer containing molybdenum vacancy, $V_{Mo}$, and vacancy complexes as $V_{Mo+3S}$ and $V_{Mo+6S}$ at both interfaces at a) Bias $= 0.00 \,V$ and b) Bias $= 1.40 \,V$. Energies are shifted with respect to their corresponding Fermi energy. The insets show the change in the electronic transmission channels at the top of the valance band. c) I-V characteristics for the same devices. The inset shows the current around the threshold voltage.}
\label{fgr:IV_trans_mo}
\end{figure}

%We then turn to the calculation of 
We calculate the electronic and transport properties of the T-H heterophase junction when molybdenum vacancy, $V_{Mo}$, and vacancy complexes $V_{Mo+3S}$ and $V_{Mo+6S}$ are present at the interface.
The LDOS of the interface is shown in Fig. \ref{fgr:ldos_v_mo} at Bias $= 0.00 \,V$ and Bias $= 1.40 \,V$. Here, the applied bias has shifted the energies. The electronic structure of the interfaces with a single Mo vacancy varies more than that of a single sulfur vacancy. In the case of larger point defects, i.e. $V_{Mo+6S}$, the electronic  structure shows several resonant states around the Fermi level which are mainly formed by Mo $4d$ states.
The defect-induced changes in the electronic structure affect the carrier injection through the junction. The transmission function (Fig. \ref{fgr:IV_trans_mo}b\&c) at the top of the valance band shows a significant enhancement when vacancies are introduced into the interfaces. Accordingly, the current is increased by up to three orders of magnitude in comparison to the perfect interface. This is due to an enhancement of carrier occupations near the Fermi level, which leads to an increase in the transmission spectrum. As a result, the interfaces with $V_{Mo+6S}$ vacancy demonstrate a threshold voltage of ≈0.5 V, half of that for a perfect interface. 

To further investigate the transport behaviour at the interface, we also plot the vector current for perfect systems and devices with $V_{Mo}$ at both interfaces (Fig. S4). Vector current displays the direction and the amount of current, coming from the left or right electrode, projected on each atom and at a specific energy channel. While the perfect interface shows dominant current scattering at the T-H boundary for low bias voltages, e.g. V=-0.5 V, the currents are delocalized in the channel region with $V_{Mo}$ at the interface, suggesting that electrons have been well transmitted from electrodes to the channel region.

%We also plot the vector current for perfect systems and devices with $V_{Mo}$ at both interfaces, \textbf{as shown in Fig. S4 in supplementary information}.
%It can be clearly seen that the presence of the vacancy at the interfaces opens up further energy channels for the charge carriers to move. 

We have also studied the influence of defect concentrations on the transport properties through the interfaces. Here, , we fixed the length of the channel but varied its width, including the interfaces with a single $V_{Mo}$.
Fig. \ref{fgr:IV_concen} shows the difference in the conductance through devices ($\Delta G \,=\, G_{defect} - G_{perfect}$) as a function of the devices' areas. It is evident that the conductance reduces with decreasing the defect concentration approaching the value of the perfect interface for zero-defect density. Variation of channel widths has two different effects on the transport properties: On one hand, the number of transport channels increases with the width of the channel. On the other hand, for a constant number of defect sites, increasing the channel width leads to a decrease in the carrier densities around the Fermi Level. The former effect is canceled out by subtracting the conductivity of the system from the corresponding one with the pristine interface. As a result, for wider channels (lower concentrations), the increase in electrical conductivity is linearly reduced.

\begin{figure*}[!htb]
  \begin{minipage}[c]{0.50\textwidth}
    \includegraphics[width=1.00\textwidth]{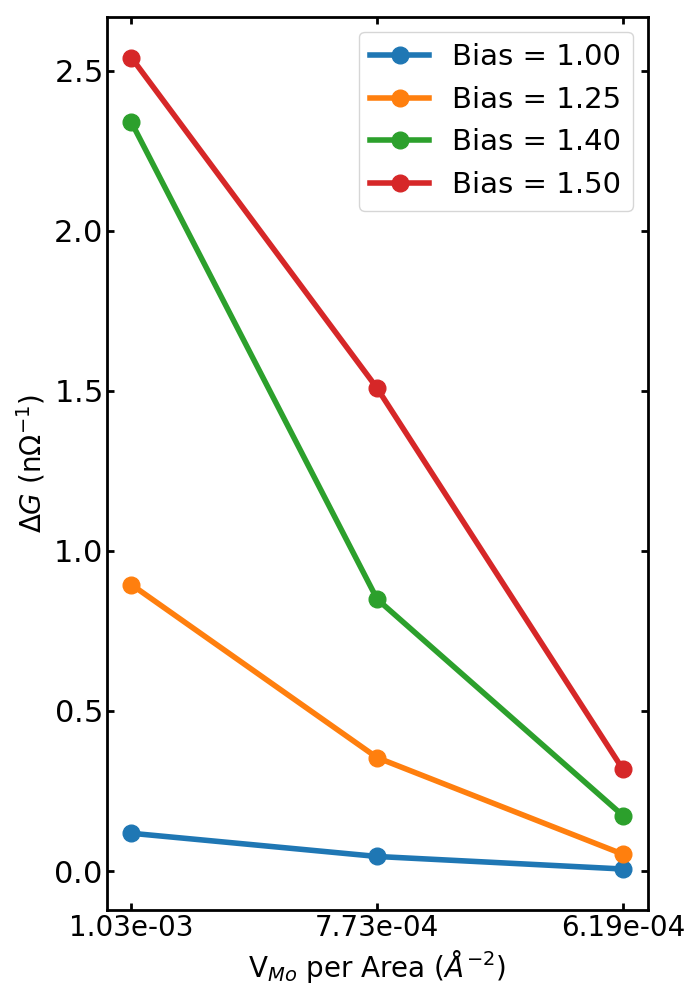}
  \end{minipage}\hfill
  \begin{minipage}[c]{0.45\textwidth} 
    \caption{(Color online) Difference in the conductance through perfect systems and device with $V_{Mo}$ ($\Delta G \,=\, G_{defect} - G_{perfect}$) as function of the device's area at different bias voltages. Since number of vacancies is the same for all cases, it is also a function of the defect density. It is true that more transport channels are added to wider devices, however, at the same time defect density is reduced, hence, decreasing the current through defective devices. Thus, conductance difference is approaching to zero, as the system gets wider.}
	\label{fig:IV_concen}
  \end{minipage}
\end{figure*}

%\newpage
% ---------------------------------------------
% ---------------------------------------------
\subsection{Antisites}
\begin{figure}[!htb]
\centering
\includegraphics[width=0.85\textwidth]{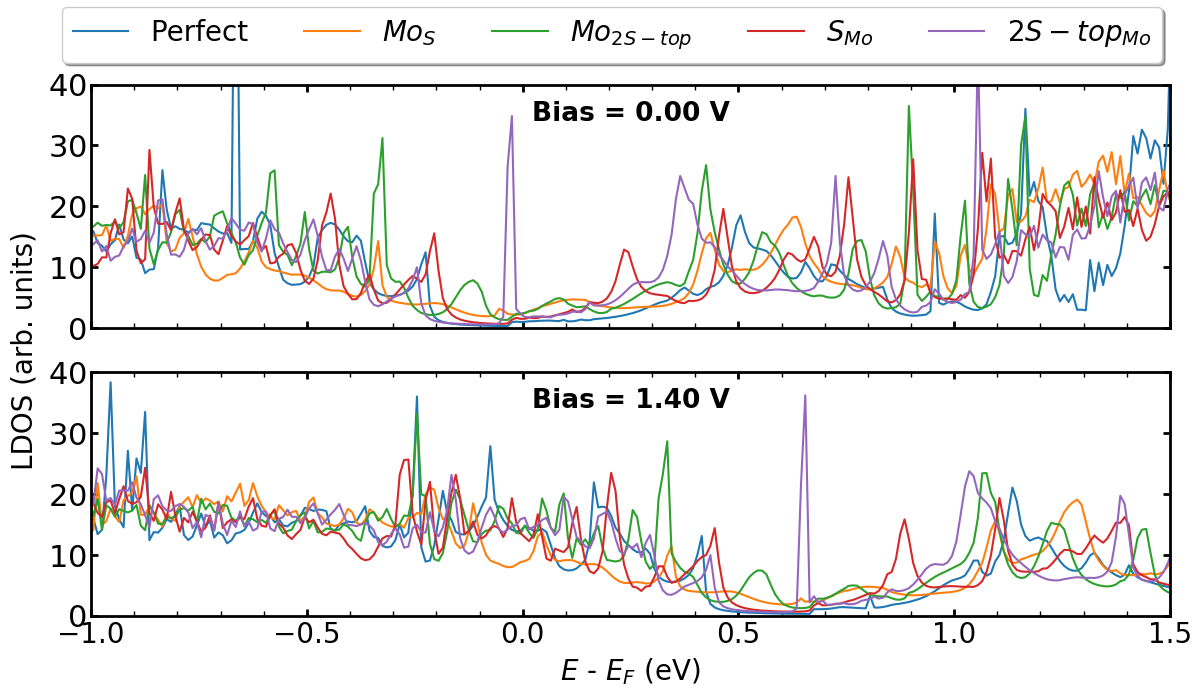}
\caption{(Color online) LDOS at the left interface of T-H heterophase junction containing various substitutions; $Mo_{S}$, $Mo_{2S-top}$, $S_{Mo}$, ${2S-top}_{Mo}$. Energies are shifted with respect to their corresponding Fermi energy.}
\label{fgr:ldos_v_antisites}
\end{figure}

Because vacancy and antisite are both high electron-scattering centers, their presence in TMDs can impair sample mobility \cite{Hong2015}. We further investigate the influence of antisites defects, such as $Mo_{S}$, $Mo_{2S-top}$, $S_{Mo}$, and ${2S-top}_{Mo}$, at the interfaces of T-H heterophase junctions. Among all the considered antisites, the situation where molybdenum vacancy is substituted with two sulfurs (${2S-top}_{Mo}$) provides the most pronounced defect associated states at the valence band edge (Fig. \ref{fgr:ldos_v_antisites}). When compared with the Mo vacancy, the defect states are more localized and originated mainly from hybridization between the $Mo_{d}$-$S_{p}$ orbitals. The contributing orbitals to the LDOS at the left interface are shown in Fig. S5 in the supplementary information.

\begin{figure}[!htb]
\centering
\includegraphics[width=1.00\textwidth]{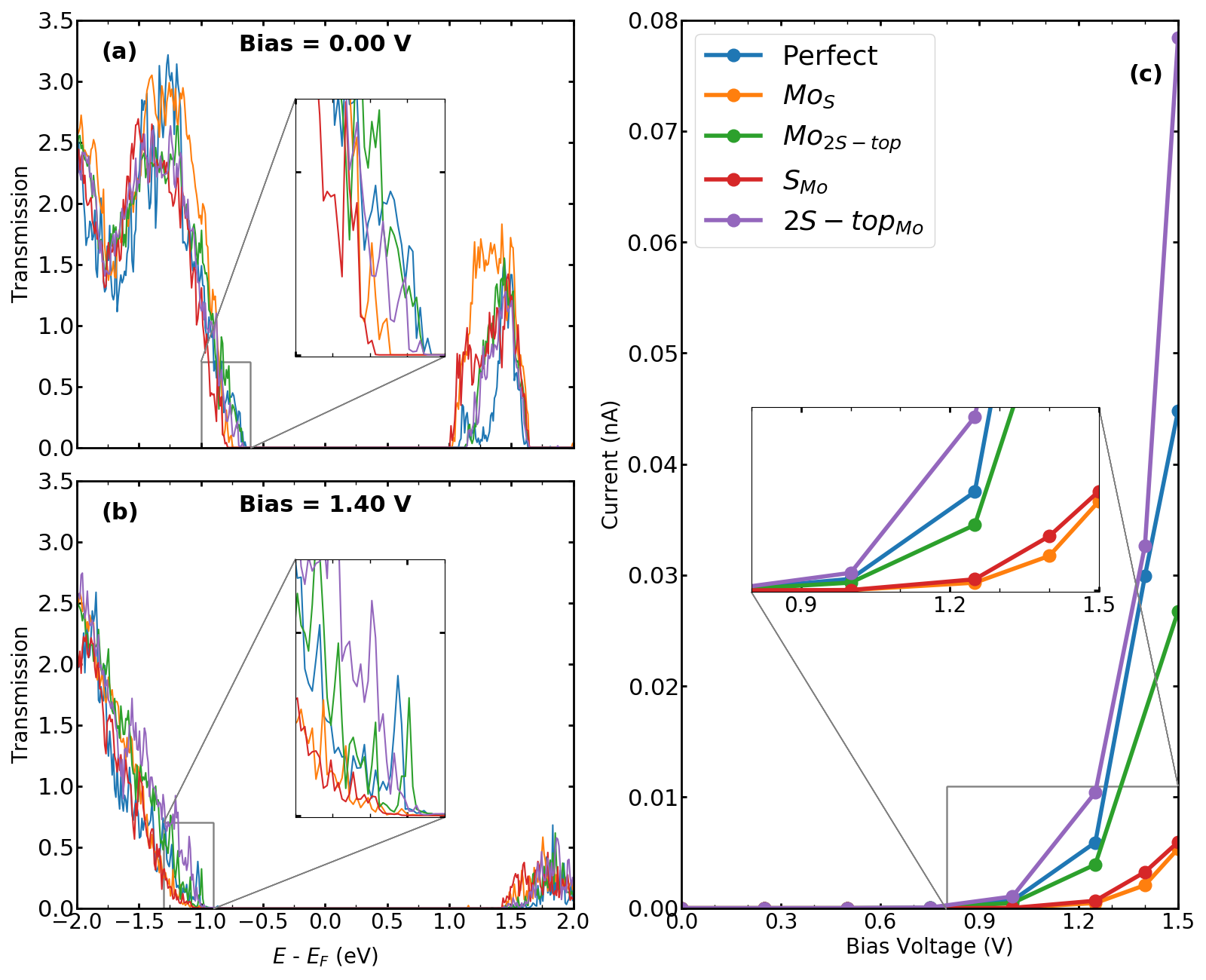}
\caption{(Color online) Transmission spectra for T-H heterophase junction of MoS$_2$ monolayer containing various substitutions; $Mo_{S}$, $Mo_{2S-top}$, $S_{Mo}$, ${2S-top}_{Mo}$, at both interfaces at a) Bias $= 0.00 \,V$ and b) Bias $= 1.40 \,V$. Energies are shifted with respect to their corresponding Fermi energy. The insets show the change in the electronic transmission channels at the top of the valance band. c) I-V characteristics for the same devices. The inset shows the current around the threshold voltage.} \label{fgr:IV_trans_antisites}
\end{figure}

In Fig. \ref{fgr:IV_trans_antisites}a,b, transmission spectra for phase-engineered devices based on MLs~MoS$_2$, containing various substitutions, are displayed at Bias $= 0.00 \,\&\, 1.40 \,V$.
Fig. \ref{fgr:IV_trans_antisites}c shows the corresponding IV characteristics as a function of bias voltages up to $1.50 V$. When vacancies are substituted with sulfur or molybdenum atoms, the current stays in the same order as for the device with perfect interfaces. Only for the case of ${2S-top}_{Mo}$, current has been slightly increased due to the presence of midgap defect states observed in LDOS and the enhancement in the transmission probabilities. It's also worth noting that, in contrast to the case of perfect interfaces, the presence of antisite defects inevitably results in more phonon scattering channels, which may be beneficial in lowering lattice thermal conductivity.

\newpage
%---------------------------------------------
% Conclusion
% ---------------------------------------------
\section{Conclusion} \label{sec:con}
In the present paper, transport properties of charge carriers through devices based on metallic (1T) and Semiconductor (2H) phases of MoS$_2$ monolayers are investigated. Various point defects are present at both interfaces: $V_{S}$, $V_{2S-top}$, $V_{2S-par}$, $V_{Mo}$, $V_{Mo+3S}$, $V_{Mo+6S}$, $Mo_{S}$, $Mo_{2S-top}$, $S_{Mo}$, and ${2S-top}_{Mo}$. The first-principles simulations and NEGF technique are used to compute the LDOS, transmission curves, and IV characteristics of perfect and defective devices under bias in the range of $0.00 \,V$ till $1.50 \,V$. Our systematic study shows that defects at the interfaces provide the opportunity for further improvement of the transport properties of such devices. 
More notably, we found that transport properties are enhanced in the presence of energetically favorable intrinsic point defects. In contrast to the scattering character of defects in 2H-phase MoS$_2$, at the interface, they lead to emergence of resonant states close to the Fermi level, thereby giving rise to the enhancement of the current flow. In particular, creating a molybdenum vacancy induces defect midgap states in the LDOS and improves the transport characteristics, which, in turn, leads to an increase in the current up to two orders of magnitude. The knowledge developed in this study could pave the way for the promising applications of lateral heterojunctions of 1T/2H~MoS$_2$ monolayers in field effect devices.

%---------------------------------------------
% Acknowledgments
% ---------------------------------------------
\section*{Author Contributions}
Conceptualization, M.B. and M.G.-A; investigation, M.B.; methodology, M. B., and M.G.-A.; writing, M.B., M.G.-A., T. F.; supervision, M. G.-A, T. F. All authors have read and agreed to the published version of the manuscript.

\section*{Conflicts of interest}
There are no conflicts to declare.

\section*{Acknowledgements}
We thank the DFG funded research training group "GRK 2247".M.B. acknowledges the support provided by DAAD and the PIP program at Bremen university. M.B. also thanks Dr. Nick Papior for his support during the transport calculations and Dr. Miguel Pruneda for his help to produce well-performed basis-sets and pseudopotentials. 

%\end{acknowledgments}

%---------------------------------------------
% References
% ---------------------------------------------
%\section{References}

\bibliographystyle{unsrtnat}
\bibliography{1t_2h_device}

%\newpage
%---------------------------------------------
% Supplementary Information
% ---------------------------------------------
%\section{Supplementaryinfo}
%\import{supplementary-info/}{supplementaryinfo.tex}
%\onecolumngrid

%\externaldocument[SI]{supplementary-info/supplementaryinfo}
%\include{supplementary-info/supplementaryinfo}

\end{document}